\documentclass[pra, twocolumn]{revtex4-2}
\usepackage{amsmath,amsthm,amssymb}
\usepackage{mathtext}
\usepackage[T1,T2A]{fontenc}
\usepackage[utf8]{inputenc}
\usepackage[english]{babel}
\usepackage{graphicx}
\usepackage[usenames]{color}
\usepackage{colortbl}

\begin{document}



\title{Theory of nonlinear whispering gallery mode dynamics in a cylindrical microresonator with a radius variation}


\author{Alena Yu. Kolesnikova}
\email{a.kolesnikova@g.nsu.ru}
\author{Ilya D. Vatnik}

\affiliation{Novosibirsk State University, 2 Pirogova Street, Novosibirsk 630090, Russia}


\date{\today}

\begin{abstract}
We propose a comprehensive model describing the Kerr nonlinear dynamics of an electric field in a cylindrical microresonator with an effective radius variation, coupled to a radiation source. The proposed system of equations for coupled azimuthal modes takes into account full azimuthal dispersion as well as the influence of the radiation source on the field in the microresonator with the coupling coefficients determined experimentally. The model appears a powerful tool to study nonlinear effects, generation axial-azimuthal modes and optical frequency combs. We illustrate the power of the model with optimization of the coupling point of the light source, getting two order of magnitude improvement for the nonlinear threshold. 
\end{abstract}


\maketitle


\section{Introduction}

Optical microresonators are currently pushing forward many directions in photonics. Due to their small mode volume and high quality factors, microresonators are an excellent test-bed for probing nonlinear and quantum optics problems~\cite{PASQUAZI20181}. For instance, the optical frequency combs~(OFCs) generated in microcavites unlock spectroscopic devices of exceptional precision \cite{Niu2023}. Depending on the free spectral range of the comb, different applications are preferred: while high repetition rate of OFCs are appreciated in optical communications and subterahertz generation,  spectroscopy applications such as dual-comb spectroscopy may benefit from lower repetition rate OFCs \cite{Sugiyama2023}.

\begin{figure}[htbp!]
\includegraphics[width=6.5cm]{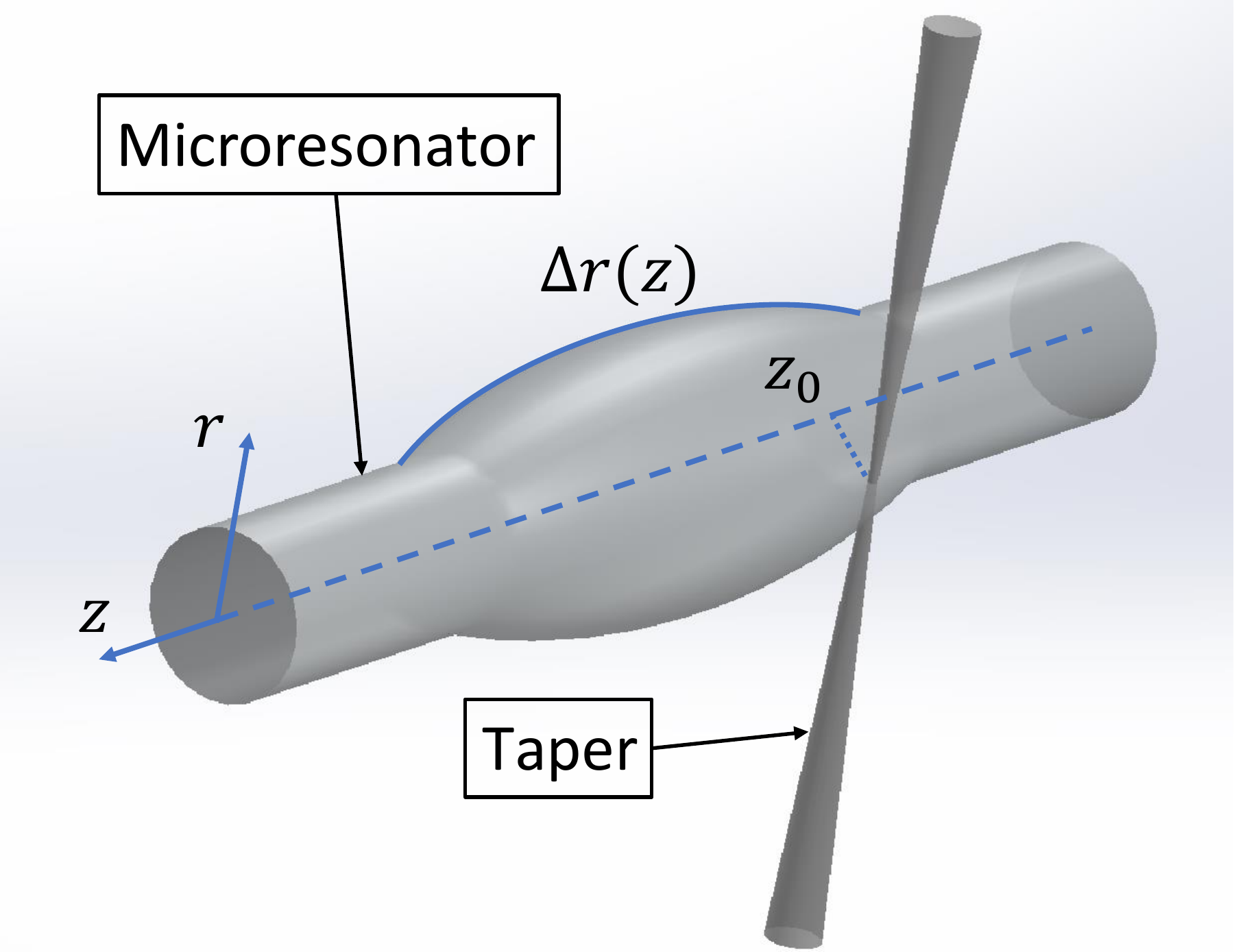}
\caption{SNAP platform: a cylindrical microresonator with an effective radius variation based on standard optical fiber coupled to an input/output radiation taper.}\label{Fig1}

\end{figure}

There is a microresonator platform possibly facilitating a low repetition rate comb generation that is called Surface Nanoscale Axial Photonics (SNAP)~\cite{Sumetsky:11}. The platform exploits a cylindrical microresonator, frequently made of standard optical fiber with removed plastic cladding, with introduced small-scale radius variations (see Fig.~\ref{Fig1}).  The effective radius variation plays the role of an optical potential that constrains  whispering gallery modes (WGM), splitting each azimuthal mode with a certain number of azimuthal nodes into a number of axial modes (with different numbers of axial nodes). The precise design of tiny variations reduces the spectral distance between adjacent axial resonances down to hundreds of megahertz \cite{Bochek:19}, paving the way to a low repetition frequency comb, while high accuracy of modifications may help to control the mode dispersion for efficient generation of the OFCs. However, no nonlinear process has yet been observed experimentally in such microresonators, as most attempts have been concentrated on bottle-like resonators \cite{Jin:21,Pollinger:10, ZhuXiaoJiangShiZhang+2019+931+940} with much larger radius variations. Those are similar to SNAP cavities but have considerable larger radius variations and thus lower mode volumes as well as much larger free spectra range for axial modes, so are not suitable for low-repetition rate OFC. That's why it's of large interest to build up a model that may help to describe nonlinear generation in SNAP cavities. 

To describe the field distribution and dynamics of WGMs in the SNAP microresonator several mathematical models may be involved. The well-established paradigm is a use of generalized Lugiato-Lefever equation \cite{lugiato2018lugiato}, that was successfully implemented for bottle-like resonators \cite{Kartashov:18,Oreshnikov:17}. The obstacle preventing use of such models for SNAP cavities is that the coupling element, launching light to the cavity, may disturb the optical potential \cite{Sumetsky:12} that is negligibly small in bottle-like resonators but should be taken into account in SNAP cavities as the optical potential in latter is smaller.  Finally, the aforementioned models do not include the coupling strength, therefore they are not suitable for determining the experimental parameters for quantitative comparison with the experiments and prediction of the thresholds.

Another approach is being developed in the frame of theory of SNAP devices. Thus,  a model based on stationary Schrödinger equation describing axial mode distribution and eigenfrequencies was proposed in \cite{Sumetsky:12}. The model is useful for designing the necessary effective radius variations. In the works \cite{Suchkov:17,Crespo-Ballesteros2022} the dynamical extension of this model was done to describe an evolution of a single azimuthal whispering gallery mode, or in the further work  \cite{Crespo-Ballesteros2021} for two azimuthal modes, though it does not account for arbitrary number of azimuthal modes as well as for experimentally determined setup parameters. 

Finally, in our recent work \cite{Kolesnikova:22} we presented the mathematical model of nonlinear dynamics of axial-azimuthal modes in the microresonator coupled to a source. But this model does not comprise the real experimental parameters, thus not allowing to judge the feasibility of the optical frequency comb in the experiment.


In this work, we present a system of equations derived from first principles, that is a most complete generalization of all existing models for describing the SNAP system. The dynamical model includes the nonlinear interaction of azimuthal modes, the total dispersion of axial-azimuthal modes (including material), and the influence of the radiation source on the propagation of modes in the microcavity. Using the proposed model, we expose the importance of impact made by the coupling element on the nonlinear threshold and demonstrate that large axial extension of SNAP modes allows optimization of coupling point, reducing the threshold from a hundred of watts to experimentally achievable values of the order of hundreds of milliwatts.

\section{Derivation of mathematical model}

\subsection{Derivation of a system of nonlinear equations for the dynamics of azimuthal modes}\label{sec:derivation}

To derive the model, we start from the Maxwell's equations and obtain the wave equation under the condition of propagation of radiation in a homogeneous isotropic dielectric matter (for example, fused silica). We neglect \((\vec \nabla \vec E)\) due to the smallness of the nonlinearity and consider only a linear polarization:
\begin{multline}
     \Delta E(\vec r,\omega) + n(\omega)^2\cfrac{\omega^2}{c^2} E(\vec r,\omega) = \\=-\mu_0\omega^2\left(P_{NL}(\vec r,\omega)+P_{p}(\vec r,\omega)\right).\label{Eq_Wave}
\end{multline}
Starting from Eqs.(\ref{Eq_Wave}), we sequentially derive a linear dynamics equation for WGMs, then take into account the radiation source, and finally Kerr nonlinear terms.

\subsubsection{Modes of the infinite cylinder}

At the first stage, we found the modes of an infinite cylinder with no radius variations \(\Delta r =0\). The electric field can be represented as \(E(\vec r,\omega)=R(\sqrt{k^2-\beta^2}r)e^{im\varphi}e^{i\beta z}\), where \(k=\omega n/c\). 
We solve the following equation:

\begin{equation}
     \left(\Delta_T +\left(k^2-\beta^2\right)\right)R(\sqrt{k^2-\beta^2}r)e^{im\varphi} = 0.\label{Eq_Wave_2}
\end{equation}

In the case of \(\beta=0\), the solution of \eqref{Eq_Wave_2} was found, which corresponds to a distribution of the azimuthal-radial infinite cylinder mode \(e_{m,p}(r,\varphi)\equiv R(k_{m,p}r)e^{im\varphi}=Ai\left(-(2/m)^{1/3}\left(T_{m,p}r/\tilde a-m\right)\right)e^{im\varphi}\), where \(Ai(x)\) - the Airy function, \(T_{m,p}\) - the \(p\)-th zero of the \(m\)-th Bessel function, \(\tilde a=r_0+P/\gamma\), \(\gamma=\sqrt{n^2-1}\omega_{m,p}/c\), \(P=1\) for TE-modes and \(P=1/n^2\) for TM-modes~\cite{Demchenko:13}.

Here \(k_{m,p}=\omega_{m,p} n(\omega_{m,p})/c\), \(\omega_{m,p}\) is the frequency of the azimuthal-radial mode of an ideal cylindrical microresonator with no radius variations, \(m\) and \(p\) are the azimuthal and radial quantum numbers respectively. \(\omega_{m,p}\) is determined from the characteristic equation (A3) in ~\cite{Demchenko:13}. It is noteworthy that the expression for resonant frequencies \(\omega_{m,p}\) may  not only take into account the geometric mode dispersion as in ~\cite{Demchenko:13}, but also comprise the material dispersion, if we assume \(n\equiv n(\omega_{m,p})\) and solve the implicit equation (A3).

In further considerations we analyze only one radial mode with \(p=1\), since it is the easiest to excite, and we further omit the index \(p\) for simplicity. Generally, the proposed model can be easily extended to take into account other radial modes.


\subsubsection{Stationary model for a single azimuthal mode}
\label{sec:stat_model}

At the second stage, we take into account a radius variation and find the stationary equation for a single azimuthal-radial mode represented  in the following form:

\begin{equation}
    E(\vec r)=A_m(z, \omega)R(\sqrt{k^2-\beta^2}r)e^{im\varphi}.\label{Eq3}
\end{equation}

The expansion (\ref{Eq3}) is valid for small radius variations \(\Delta r(z)\): \(\Delta r(z)\ll r_0\), where  \(r_0\) is an undisturbed radius of the cylinder. We substituted the expression for the field into \eqref{Eq_Wave} in linear regime and divided the equation into axial and transverse parts:

\begin{multline}
     \left(\Delta_T +\left(k^2-\beta^2\right)\right)R(\sqrt{k^2-\beta^2}r)e^{im\varphi}A_m(z,\omega)+\\ + \left(\cfrac{\partial^2}{\partial z^2} +\beta^2\right)A_m(z,\omega)R(\sqrt{k^2-\beta^2}r)e^{im\varphi}= 0.\label{Eq_Wave_3}
\end{multline}



In the first-order approximation for the amplitude  \(A_m(z)\), in the presence of an effective radius variation for \(m \gg 1\) the wave vector has a small component \(\beta\) directed along the axis \(z\), such that  \(\beta \ll k_{m}\). In this case, we can assume that the characteristic equation for frequencies changed up to replacement \(\sqrt{k^2-\beta^2} \to k_{m}\) and \(\beta\) can be found from following expression~\cite{Sumetsky:11}: 

\begin{equation}
\cfrac{\omega_{m}n(\omega_{m}) r_0}{c}=(r_0+\Delta r(z))\sqrt{\cfrac{\omega^2n(\omega)^2}{c^2}-\beta^2}.\label{Eq6}
\end{equation}

We expand \(\omega^2n^2(\omega)\) near \(\omega_{m}\) and \(r_0\) to the first order, taking into account the dependence of the refractive index \(n(\omega)\equiv n(\omega,r(z))\) on the coordinate resulting from the introduction of a radius variation:

\begin{equation}
\beta^2=2k_{m}^2\left(\cfrac{\Delta r(z)}{r_0}+\cfrac{\Delta n(z)}{n_m}+K_m\cfrac{\Delta\omega_m}{\omega_m}\right).\label{Eq_beta}
\end{equation}
Here \(k_m=\cfrac{\omega_m n(\omega_m)}{c}\), \(K_m = 1+\cfrac{\omega_m}{n_m}\cfrac{\partial n}{\partial \omega}(\omega_m)\) is the coefficient of material dispersion, \(\Delta \omega_m = \omega-\omega_m\), \(\Delta r(z)\) and \(\Delta n(z)\) determine the effective radius variation. 

In this case, the transverse parts in the equation \eqref{Eq_Wave_3} is equal to the zero, according to the equation \eqref{Eq_Wave_2}, and we receive the equation for a slowly varying amplitude of the azimuthal mode:
\begin{equation}
         \cfrac{\partial^2A_m(z,\omega)}{\partial z^2} +\beta^2
     A_m(z,\omega)= 0.\label{Eq5}
\end{equation}

We then substitute  \(\beta\) in  the equation \eqref{Eq5} with \eqref{Eq_beta} and obtain the Schrödinger equation describing the stationary distribution of the eigenmodes of the resonator axial-azimuthal modes and resonance frequencies. 

\begin{equation}
     \cfrac{\partial^2A_m(z,\omega)}{\partial z^2} +V_m(z)A_m(z,\omega)= E_mA_m(z,\omega).\label{Eq_Shrod_stat}
\end{equation}

Here the potential is determined by the effective radius variation \(V_m(z)=2k^2_m\cfrac{\Delta r_{eff}(z)}{r_{0_{eff}}}\), where \(\cfrac{\Delta r_{eff}(z)}{r_{0_{eff}}}=\cfrac{\Delta r(z)}{r_0}+\cfrac{\Delta n(z)}{n_m}\). The resonant frequencies correspond to the energy levels in the potential and are related as follows: \(E_m = -2k^2_m\cfrac{\Delta\omega_m}{\omega_m}K_m\). 

The equation \eqref{Eq_Shrod_stat} is virtually the same with obtained in \cite{Sumetsky:11}. The feature of the current version of the equation is taking into account the material dispersion of azimuthal modes through the coefficient \(K_m\). Despite the smallness of the correction \(K_m -1\ll 1\), it might be crucial for considering the interactions between different azimuthal modes, as it imposes additional azimuthal dispersion as well as dissimilar axial free spectral ranges for different azimuthal modes for the same $\Delta r_{eff}(z)$.

In order to take account for the internal losses in the microresonator medium, one can modify the energy definition as follows:
\(E_m = -2k^2_m\cfrac{\Delta\omega_m}{\omega_m}K_m+i\Gamma\). Here losses $\Gamma$ may in principle depend on azimuthal or radial quantum numbers, and are defined by the intrinsic losses within the cylinder media as well as by the surface quality.


\subsubsection{Dynamical model with a source and nonlinearity}
To generalize the approach presented in the previous section for the arbitrary number of azimuthal modes, the field should be represented as 
\begin{equation}
    \begin{array}{c}
    E(\vec r,t)=\sum\limits_{m}A_{m}(z,t)\exp{(i\omega_{m}t})e_{m}(r,\varphi)+c.c.,\\
    E(\vec r,\omega)=\sum\limits_{m}A_{m}(z,\Delta\omega_{m})e_{m}(r,\varphi)+c.c.
    \end{array}\label{Eq_field}
\end{equation}
with normalization $\max |e_m(r, \varphi)|=1$.
Substituting the field in the form of \eqref{Eq_field} into the equation \eqref{Eq_Wave}, and taking into account the equations \eqref{Eq_Wave_2} and \eqref{Eq_beta}, one obtains:



\begin{equation}
\begin{gathered}
     \sum\limits_m\Bigg(\cfrac{\partial^2}{\partial z^2} + 2k_m^2\Bigg(\cfrac{\Delta r_{eff}(z)+\Delta r_t(\vec r)}{r_{0_{eff}}}+K_{m}\cfrac{\Delta \omega_m}{\omega_m}\Bigg)-\\-i\Gamma \Bigg)A_me_{m}=  -\cfrac{\chi^{(3)}\omega^2}{c^2}E^3-\cfrac{\chi\omega^2}{c^2}E_p.\label{Eq9}
\end{gathered} 
\end{equation}

Here we consider the Kerr nonlinearity $P_{NL}=\epsilon_0\chi^{(3)}E^3$, as the SNAP cavities usually assumed to be on silica. The pump field $E_p(\vec r, t)=E_p(\vec r)e^ {i\omega_{p}t}$, where $\omega_p$ is the pump frequency.

Importantly, we here introduce an additional term \(\Delta r_t(\vec r)\) to stress out the additional effective radius variation  that may appear in the presence of a coupling element, for instance, a taper being in contact with the cavity \cite{Sumetsky:12,Vitullo:20}. Complex effective radius variation \(\Delta r_t(\vec r)\)=\(\Delta r'_t(\vec r)-i\Delta r''_t(\vec r)\) accounts for both the change in the structure of the microcavity eigenmodes, as well as for additional losses.

To pass to the dynamic equation, we take the inverse Fourier of \eqref{Eq9} and put out of brackets the coefficient \(2k^2_mK_m/\omega_m\) and taking the inverse Fourier transform:

\begin{equation}
\begin{gathered}
     \sum\limits_m2\omega_mK_mn^2_m\Bigg(i\cfrac{\partial }{\partial t}-\cfrac{\omega_m}{2k_m^2K_m}\cfrac{\partial^2}{\partial z^2}- \cfrac{\omega_m}{K_m}\cfrac{\Delta r_{eff}(z)}{r_{0_{eff}}}-\\-\cfrac{\omega_m}{K_m}\cfrac{\Delta r_t(\vec r)}{r_{0_{eff}}}+i\Gamma\Bigg)A_m(z,t)e_{m} 
     = -e^{-i\omega_mt}\chi^{(3)}\cfrac{\partial^2 E^3}{\partial t^2}+\\+e^{i\Delta\omega_{p}t}\chi\omega_p^2E_p(\vec r),\label{Eq10}
\end{gathered} 
\end{equation}
where \(\Delta\omega_p = \omega_p-\omega_m\).
Multiplying the equation \eqref{Eq10} by \(e^*_{m}\) and integrating over the cross-section of the cylinder, implying that modes with different azimuthal numbers are orthogonal, we obtain:

\begin{equation}
\begin{gathered}
     i\cfrac{\partial A_m }{\partial t}-\cfrac{\omega_m}{2k_m^2K_m}\cfrac{\partial^2A_m }{\partial z^2} - \cfrac{\omega_m}{K_m}\cfrac{\Delta r_{eff}(z)}{r_{0_{eff}}}A_m +i\Gamma A_m +\\
     +e^{-i\omega_mt}\cfrac{\chi^{(3)}\int e^*_{m}\cfrac{\partial^2 E^3}{\partial t^2}d^2r}{K_m2\omega_m n_m^2S^{(eff)}_{m}}+D_mf_p(z)A_m=\\=\sqrt{\cfrac{P_{in}}{\varepsilon_0 n^2_m S^{(eff)}_{m}}} C_mf_p(z)e^{i\Delta\omega_{p}t}.\label{Eq_system_with_nonlinear_integral}
\end{gathered} 
\end{equation}

Here  \(f_p(z)\) is the normalized spatial distribution of the source radiation ($\int f_p(z)dz=1$ ), that for the case of the thin taper can be considered as delta-shaped: $\delta(z-z_0)$, where $z_0$ is a coordinate  of the contact between the taper and the cavity. 
 
\begin{equation}
 \begin{gathered}
     C_m = \cfrac{ \chi\omega_p^2}{K_m 2\omega_mn_m}\sqrt{\cfrac{\varepsilon_0}{P_{in}S^{(eff)}_m}}\int E_p(r,\varphi) e^*_m d^2r,\\
     D_m = -\cfrac{\omega_m}{K_m r_{0_{eff}}}\int \Delta r_t(r,\varphi) e^*_m d^2r \label{Eq_C_D_def}
\end{gathered} 
\end{equation}
are the coupling parameters determined by the overlap integral of the guided mode of a taper and transverse distribution of the whispering gallery mode at the cross-section of the cylinder. \(C_m\) contains information about the pump power and coupling strength. The real part of \(D_m\) contributes to the effective radius variation, and the imaginary part contributes to the losses introduced to the mode by the taper.  Note that coefficients $C_m,D_m$ have different dimension comparing to those introduced in the stationary model in \cite{Sumetsky:12,Vitullo:20}. \(S^{(eff)}_{m}=\int |e_m(r,\varphi)|^2d^2r\) is the effective mode area. Importantly, the parameters \(C_m\) and \(D_m\) are not dependent on the position of the coupling element $z_0$.

Finally, with the simplification of the nonlinear term described in the appendix \ref{app:nonlinear_term},  the equation \eqref{Eq10} is rewritten in the following form:

\begin{multline}
     i\cfrac{\partial A_m }{\partial t}-\cfrac{\omega_m}{2k_m^2K_m}\cfrac{\partial^2A_m }{\partial z^2} - \cfrac{\omega_m}{K_m}\cfrac{\Delta r_{eff}(z)}{r_0}A_m+i\Gamma A_m-\\-\frac{3\omega_{m}\chi^{(3)}}{K_m2n^2_mS^{(eff)}_{m}}F_m(\vec{A}) +D_mf_p(z)A_m\\=\sqrt{\cfrac{P_{in}}{\varepsilon_0 n^2_m S^{(eff)}_{m}}} C_mf_p(z)e^{i\Delta\omega_{p}t}, \label{Eq_system}
\end{multline}
where \(F_m(\vec A)\) is determined by the equation \eqref{Eq_F_def}.

The system of equations \eqref{Eq_system} is the most complete model that describes the dynamics of interacting azimuthal modes in a cylindrical microcavity coupled to an exciting element. To use the model, coupling parameters \(D_m\) and \(C_m\)  should be specified. Calculating the integrals in equations \eqref{Eq_C_D_def} in the general case might be meaningless, as the mode distribution within a coupling element usually is not controlled precisely. Therefore, it's necessary to relate  \(D_m\), \(C_m\) with quantities determined in the experiment, that will be done in the next section.

\subsection{Determining the coupling parameters $C_m$, $D_m$ }\label{sec:Coupling_parameters}

To establish links between the $C_m$ and $D_m$ and measurable values,  one may deduce the transmission spectrum $T(\lambda)$ of the microresonator-taper system, that is the experimentally observed function, from the system \eqref{Eq_system}. For this, one can reduce the model \eqref{Eq_system} to the simple equation for coupling between an exciting element waveguide mode and a particular whispering gallery mode. Within this simplified model, the transmission spectrum $T(\lambda,\delta_0,\delta_c)$ is defined, giving a way to gather the coupling strength coefficient  $\delta_c$ and the losses $\delta_0$ experienced by the whispering gallery mode in an experiment (see appendix \ref{app:coupled_mode_equation}). 

Shrinking \eqref{Eq_system} to the form of the equation \eqref{Eq_simple_coupled_model} may be done for a linear case, since  measurements of the transmission spectrum are performed at low powers. 
If a single azimuthal-axial mode with the azimuthal number $m$ and axial number $q$ is excited,  the amplitude is represented in the following form \(A_m(z,t)=a_m(t)e^{i(\omega_p-\omega_{m})t}Z_{q}(z)\) with normalization \(\max Z_q(z)=1\)
, and the dynamical equation is reduced to

\begin{multline}
     i\cfrac{\partial a_m}{\partial t}Z_q-\Delta \omega_{p} a_mZ_q-\cfrac{\omega_m}{2k_m^2K_m}\cfrac{\partial^2 Z_q}{\partial z^2}a_m - \\- \cfrac{\omega_m}{K_m}\cfrac{\Delta r_{eff}(z)}{r_{0_{eff}}}a_mZ_q +i\Gamma Z_qa_m+D_mf_p(z)Z_qa_m=\\=\sqrt{\cfrac{P_{in}}{\varepsilon_0 n^2_m S^{(eff)}_{m}}} C_mf_p(z).\label{Eq16}
\end{multline}

Multiplying the equation \eqref{Eq16} by \(Z_{q}(z)\) and integrating over \(z\):

\begin{multline}
i \cfrac{\partial a_m}{\partial t}L^{(eff)}_q-\Delta\omega_p L^{(eff)}_qa_m+D_mZ_q^2(z_0)a_m-\\-\int\left(Z_q\cfrac{\omega_m}{2k_m^2K_m}\cfrac{\partial^2 Z_q}{\partial z^2} + Z_q\cfrac{\omega_m}{K_m}\cfrac{\Delta r_{eff}(z)}{r_{0_{eff}}}Z_q\right)dz a_m+\\+i\Gamma L^{(eff)}_qa_m= \sqrt{\cfrac{P_{in}}{\varepsilon_0 n^2_m S^{(eff)}_{m}}} C_m
Z_q(z_0).\label{Eq17}
\end{multline}
Here $L^{(eff)}_q=\int Z_q^2(z) dz$ - the effective mode length.

The term \(\int\left(Z_q\cfrac{\omega_m}{2k_m^2K_m}\cfrac{\partial^2 Z_q}{\partial z^2} + Z_q\cfrac{\omega_m}{K_m}\cfrac{\Delta r_{eff}(z)}{r_{0_{eff}}}Z_q\right)dz \) can be represented as \(\langle q\vert\hat H\vert q\rangle \), where \(\hat H = \cfrac{\omega_m}{2k^2_mK_m}\left(\cfrac{\partial ^2}{\partial z^2} + 2k^2_m\cfrac{\Delta r_{eff}(z)}{r_{0_{eff}}}\right)\). According to the stationary Schrödinger equation \eqref{Eq_Shrod_stat}, this matrix element expresses the energy of the mode with the number \(q\):

\begin{equation}
    \langle q\vert\hat H\vert q\rangle=E_q L^{(eff)}_q=(\omega_{m}-\omega_{m,q})L^{(eff)}_q.\label{Eq18}
\end{equation}

Thus, redefining the source detuning $\Delta\omega_p=\omega_p-\omega_{m,q}-\Omega$ ($\Omega=\operatorname{Re}( D_m)Z_q^2(z_0)/L^{(eff)}_q$), we get

\begin{multline}
i \cfrac{\partial a_m}{\partial t}-\Delta\omega_p a_m+i\Gamma a_m+i\operatorname{Im}(D_m)Z_q^2(z_0)/L^{(eff)}_qa_m=\\= \sqrt{\cfrac{P_{in}}{\varepsilon_0 n^2_m S^{(eff)}_{m}}} C_m
Z_q(z_0)/L^{(eff)}_q.\label{Eq_reduced_simple_coupled_model}
\end{multline}

Equaling each term in the equations \eqref{Eq_simple_coupled_model} and \eqref{Eq_reduced_simple_coupled_model}, we get the relationship between $C_m,D_m$ and experimentally observable $\delta_0,\delta_c$ and $\Omega$, that is the resonance frequency shift owing to additional effective radius variations introduced by the taper: 

\begin{equation}
\begin{gathered}
\operatorname{Re}(D_m) = \Omega\cfrac{L^{(eff)}_q}{Z^2_q(z_0)},\\
\operatorname{Im}(D_m) = (\delta_0+\delta_c-\Gamma)\cfrac{L^{(eff)}_q}{Z^2_q(z_0)}, \\
C_m=-i\left(\delta_c\cfrac{L_q^{(eff)}}{Z_q^2(z_0)}\right)^{\frac{1}{2}}.\label{Eq_CD_through_deltas}
\end{gathered}
\end{equation}


\subsection{Measurement of coupling parameters: an example}
\label{sec:meas_CD}
As per equation \eqref{Eq_CD_through_deltas}, to find $C_m$ and $D_m$ within the proposed model one should not only define $\delta_c,\delta_0$ from a transmission spectrum of a cavity-taper system, but also know mode intensity at the coupling point $Z_q^2(z_0)$ and the effective length $L_q^{(eff)}$ for the azimuthal-axial cavity mode under test. Fortunately, it can be derived from measurements of $\delta_c,\delta_0$ for different excitation points $z_0$. Indeed, from  equation \eqref{Eq_CD_through_deltas} follows:


\begin{equation}
\begin{gathered}
\Omega(z_0)=\operatorname{Re}(D_m) \cfrac{Z^2_q(z_0)}{L^{(eff)}_q},\\
\delta_c(z_0) = |C_m|^2\cfrac{Z^2_q(z_0)}{L^{(eff)}_q},   \\
\delta_0(z_0) = (\operatorname{Im}(D_m)- |C_m|^2)\cfrac{Z^2_q(z_0)}{L^{(eff)}_q}+\Gamma.\label{Eq_deltas_through_CD}
\end{gathered}
\end{equation}

 Interestingly, \(\delta_0\) and \(\delta_c\) depend on $z_0$ and proportional to the axial mode spatial distribution  \(Z_q^2(z)\) (Firstly it was revealed in \cite{Sumetsky:12} with emphasis on transmission spectrum properties).  The experimental dependencies \(\delta_c(z_0)\) and \(\delta_0(z_0)\) for a mode \(q\), derived from  the transmission spectrum $T(z_0,\lambda)$ at different positions of the taper along the \(z\) axis (see appendix \ref{app:coupled_mode_equation}) thus can be fitted to get $C_m$, $D_m$, as well as $\Gamma$.

We demonstrate the feasibility of the method, carrying out measurements of coupling parameters for a taper in contact with a SNAP cavity based on a piece of standard optical fiber SMF-28 with $r_0= 62.5\, \rm \mu m$. To localize modes along \(z\) axis, we introduced a bell-shaped effective radius variation $\Delta r_{eff}$ with local heating by $CO_2$ laser \cite{Sumetsky:11}.

\begin{figure}[htbp!]
\includegraphics[width=7cm]{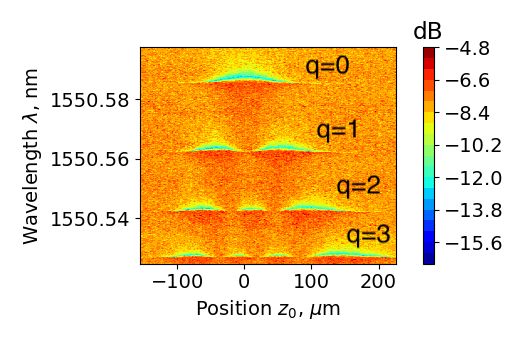}
\caption{The measured spectrogram $T(z_0,\lambda)$ of the SNAP cavity with a bell-shape effective radius variation.}\label{fig:spectrogram_all_modes}
\end{figure}

The measured spectrogram $T(z_0,\lambda)$ represents spatial distribution of axial modes with numbers \(q=0..3\) (see Fig.~\ref{fig:spectrogram_all_modes}). The spectral resolution was not worse than 5 MHz. The first consequence of the equation \eqref{Eq_deltas_through_CD} is that the linewidth of the resonance $\delta_0+\delta_c$ is maximal at the mode anti-nodes. The second is that the typical resonance width for a mode with a lower number \(q\) is larger, as the effective mode length \(L_q^{(eff)}\) is smaller.


\begin{figure}[htbp!]
\includegraphics[width=7cm]{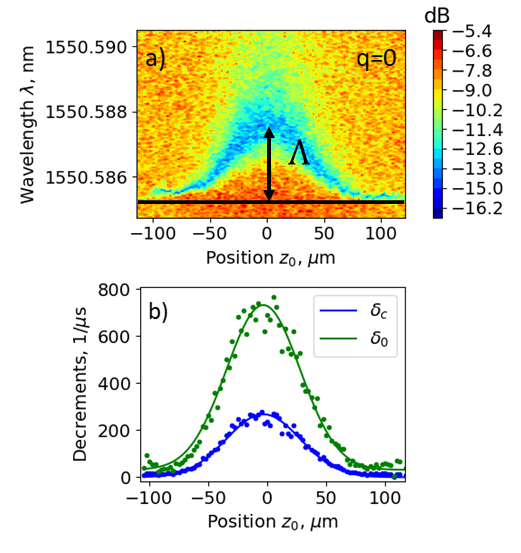}
\caption{a) The spectrogram of the SNAP system: the axial mode with \(q=0\). b) The approximation of the experimental decrements \(\delta_c\) and \(\delta_0\).}\label{fig:spectrogram_q=0}
\end{figure}

We chose the first axial mode with \(q=0\) (see Fig.~\ref{fig:spectrogram_q=0},a) to determine the coupling parameters $\delta_c(z_0), \delta_c(z_0)$. For this mode one can suppose Gaussian function for the axial distribution $Z^2(z)$,  so \(\delta_c(z)= |C_m|^2/L^{(eff)}_qe^{-(z-a)^2/w^2}\) and \(\delta_0(z)=(\operatorname{Im}(D_m) - |C_m|^2)/L^{(eff)}_qe^{-(z-a)^2/w^2} + \Gamma\) (see Fig.~\ref{fig:spectrogram_q=0},b). By fitting two lines $\delta_c(z_0),\delta_c(z_0)$ jointly with Gaussian shapes, we determined coupling parameters $\operatorname{Im}(D_m) = 7.6\cdot 10^4 \pm 1.7 \cdot 10^3 \, \rm m/s$ and $ |C_m|^2 = 2.1 \cdot 10^4 \pm 5.3 \cdot 10^2 \, \rm m/s$, and upper bound for internal losses is determined as  $\Gamma<30\, \rm \mu s^{-1}$, corresponding to the resolution of the optical spectral analyzer. 

Defining $\operatorname{Re}(D_m)$, associated with the distortion of the effective radius variation in the presence of a taper, jointly by fitting equation \eqref{Eq_deltas_through_CD} is strained because of low absolute resolution of the optical spectrum analyzer used. Nevertheless,  one can estimate the shift in the resonant frequency \(\Omega = -\Lambda\omega_m/\lambda_m=\operatorname{Re}(D_m)Z_q^2(z_0=0)/L^{(eff)}_q\) (see Fig.~\ref{fig:spectrogram_q=0},b), finding \(\operatorname{Re}(D_m) =-9.7 \cdot 10^4 \, \rm m/s \).


It should be noting that the dependence of the width and shift of axial resonances and their connection with the coupling parameters has already been studied in the work \cite{Sumetsky:12}. In this paper, the coefficients \(D\) and \(|C|^2\) are presented, which, in essence, are also coupling parameters determined by the overlap integral of the taper and resonator modes and thus proportional to the parameters \(D_m\) and \(C_m\) defined in our modes. Nevertheless, uncertainty in the normalization of the wave functions of axial modes in \cite{Sumetsky:12} does not allow finding the exact relationship between the coupling parameters and experimentally measured values.


\section{Optimizing nonlinear threshold}

Dependence of the decrements $\delta_0,\delta_c$ on the contact point along the $z$ axis is a fundamental feature of cylindrical and, in particular, SNAP microresonators, that grants the control of the loaded Q-factor. 
Note that for spherical ideal microresonators symmetry of the system imposes identical decrements for any contact point. 


Governing quality factors, in its turn, makes help in achieving nonlinear generation. From a simple coupling model, the power threshold for observing nonlinear effects may be derived using the coupling decrements \cite{Herr2012}:

\begin{equation}
    P_{in}=\cfrac{4\varepsilon_0K_m n^4_m V_{m,q}^{2^{{(eff)}}}}{3\omega_m\chi^{(3)}V_{mm,qq}}\cfrac{\delta^3}{\delta_c}.\label{Eq20}
\end{equation}

Here \(V_{mm,qq}=S_{mmmm}\int Z^4_q(z) dz\). 

Thus, in SNAP microresonators the nonlinear threshold also depends on the coupling point \(z_0\) and might be optimized. Taking into account the dependencies $P_{in} \sim \frac{(\delta_0(z_0)+\delta_c(z_0))^3}{\delta_c(z_0)} $ and equations \eqref{Eq_deltas_through_CD},  we have found that the minimum threshold is reached when coupling occurs at the point \(z_0\) defined by \(Z^2_q(z_0)=\cfrac{\Gamma L_q^{(eff)}}{2\operatorname{Im}(D_m)}\). 

The minimum power threshold is then equal to
\begin{equation}
P_{in}^{(min)}=\cfrac{9\varepsilon_0K_mn^4_m V_{m,q}^{2^{{(eff)}}}}{\omega_m\chi^{(3)}V_{mm,qq}}\cfrac{\Gamma^2\operatorname{Im}(D_m)}{ |C_m|^2}.\label{Eq_minimal_threshold}
\end{equation}

\begin{figure}[htbp!]
\includegraphics[width=7cm]{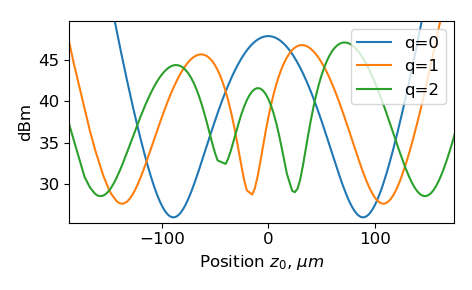}
\caption{Nonlinear threshold $P_{in}$ approximated for different coupling points $z_0$ for different axial modes.}\label{fig:thresholds_on_z}
\end{figure}

To illustrate the capabilities of the coupling optimization, we determine the minimum power threshold for the axial mode with \(q=0\) in the experimentally studied SNAP cavity (see Fig.~\ref{fig:spectrogram_q=0},a). While coupling at the maximum of the mode distribution would require\textbf{ \(P_{in}= 61 \rm\: W\)} to obtain the nonlinear threshold, the optimized coupling point yields threshold power as small as \textbf{\(P_{in}^{(min)}= 396\rm\: mW\)}. Nevertheless,  constraints may make it difficult to take advantage of such a threshold. Indeed, the point $z_0$ corresponding to the minimal power \(P_{in}^{(min)}\) is three mode widths away from the center of the mode (see Fig.~\ref{fig:thresholds_on_z}), thus earning tiny coupling parameters $\delta_c, \delta_0$ and might be hardly detectable in an experiment resonance in transmission in the case of low resolution (according to \eqref{Eq_Fano_shape}). With this, strong dependence of the threshold on $z_0$ demands the sufficient accuracy of the coupling positioning.

Generally, for each axial mode with an axial number \(q\) there are \(2(q + 1)\) points along \(z\) in which the threshold power $P_{th} (z_0)$ achieves the local minimum.  In the approximation \(\int Z_q^2(z)f_p(z)dz=Z_q^2(z_0)\) all minima are equal and are defined by the equation \eqref{Eq_minimal_threshold}. In a real system, the taper has a finite size, and the threshold value   \eqref{Eq_minimal_threshold} must be derived more accurately, giving minima at the points where the axial mode distribution function changes slowly. In other words, the minimum will be reached at the edges of the mode distributions (see Fig. \ref{fig:thresholds_on_z}). At other local minimum points the threshold is higher since the mode distribution function changes faster, and the overlap integral with the not infinitely small source is larger.

Apart from the altering the power threshold of nonlinear effects, optimization of the coupling point $z_0$ is crucial for the mode dispersion management. Indeed, according to the equation \eqref{Eq_deltas_through_CD}, there is a shift of the resonant frequency of a mode $\Omega$ proportional to the mode intensity in the coupling point $Z^2_q(z_0)$. Due to different axial distributions, axial modes with different $q$ are disturbed by the same taper in a different manner, that results in an uneven shift of the resonant frequencies and  thus additional axial mode dispersion with alternating sign. Given the experimental example under consideration with \(\operatorname{Re}(D_m) =-9.7 \cdot 10^4 \,\rm m/s \) such an alternating dispersion may be of the order of dozens of MHz for high-q modes, playing a noticeable role in the nonlinear mode generation, and should be carefully taken into account, for example, in a modulation stability analysis.

\section{Conclusion}
We present the most complete generalization of all currently existing models describing light evolution in Kerr nonlinear cylindrical microresonators with slight radius variations. The model comprises nonlinear Kerr interactions between axial-azimuthal modes and takes into account disturbances introduced by the coupling element, that may drastically change the light dynamics. We also propose a method to experimentally determine the coupling parameters. It's shown that the coupling element may introduce determinant losses to a mode, as well as additional alternating axial dispersion and thus must be taken into account while analyzing nonlinear threshold and dynamics in SNAP resonators. Within the proposed model we reveal possibilities for minimization of the nonlinear Kerr threshold by choosing the proper position of a thin taper exciting a mode. For a particular SNAP cavity made of SMF-28 fiber with mode length of 80 $\rm \mu m$  optimization decreases the threshold from 61 W down to 0.4 W.
The model may become a powerful for studying the nonlinear interactions of azimuthal-axial modes at disturbed cylinders. 


\begin{acknowledgments}
The study was supported by the Russian Science Foundation (22-12-20015), and by the Government of the Novosibirsk Region.

\end{acknowledgments}

\appendix
\section{Nonlinear term $F_m(\vec A)$}\label{app:nonlinear_term}
To obtain a system of nonlinear equations from \eqref{Eq_system_with_nonlinear_integral}, the nonlinear integral should be rewritten.
For this, the term \(E^3\) may be expanded: 
$E^3=\sum\limits_{i,j,k}\left(\tilde A_i\tilde A_j\tilde A_k+3\tilde A_i^*\tilde A_j\tilde A_k+3\tilde A_i^*\tilde A_j^*\tilde A_k+\tilde A_i^*\tilde A_j^*\tilde A_k^*\right)$.  Here \(\tilde A_i=A_{i}e^{i\omega_{i} t} e_{i}(r,\varphi)\).

Only a part of the terms constituting the sum will give a non-zero contribution to the integral $ \int e_m^{*} E^3 d^2 r$ because of oscillations in $e^{im\varphi}$:

\begin{equation}
    \int e_m(r)e_i(r)e_j(r)e_k(r)e^{i\varphi(\pm i\pm j \pm k-m)} d^2r=2\pi S_{mijk}\label{Eq_FWM}, 
\end{equation}
where \(S_{mijk}=\int e_m(r)e_i(r)e_j(r)e_k(r)rdr\). The expression \eqref{Eq_FWM} is valid for the case \(\pm i\pm j \pm k-m=0\)

Provided that the number of the azimuthal mode is \(m\gg1\) and have the same order, only terms with one conjugate amplitude remain from the entire sum: \(E^3=3\sum\limits_{i,j,k}\tilde A_i\tilde A_j\tilde A_k^*\).

We select from this sum all terms with \(\tilde A_m\):\\

\begin{enumerate}
    \item \(i=m, j=m, k=m\): \(3|\tilde A_m|^2\tilde A_m\);
    \item \(i=m, j\neq m\): \(3\sum_{\substack{j=k \\ j\neq m}} \left(\tilde A_m\tilde A_j\tilde A_k^*+\tilde A_m\tilde A_j^*\tilde A_k)\right)=6\sum\limits_{j\neq m}\left(|\tilde A_j|^2\tilde A_m\right)\). 
\end{enumerate}

Substituting the decomposition of \(E^3\) into the nonlinear integral in the equation \eqref{Eq_system_with_nonlinear_integral} we obtain:

\begin{equation}
\int e^*_{m}\cfrac{\partial^2 E^3}{\partial t^2}d^2r=-\omega_m^2F_{m}(\vec{A}),\label{Eq_non_int}
\end{equation}
where:

\begin{multline}
    F_{i}(\vec{A})=(S_{iiii}|A_{i}|^2+2\sum_{j\neq i}S_{jjii}|A_j|^2)A_{i}+\\+\frac{(\omega_i+\Delta \omega_{ijkl})^2}{\omega_i^2}\sum_{\substack{j\neq i \\ k\neq i}}S_{ijkl}A_jA_kA_l^*e^{i(\Delta \omega_{ijkl})t},\label{Eq_F_def}
\end{multline}
here \(\Delta \omega_{ijkl}=-\omega_i+\omega_j+\omega_k-\omega_l\), \(l=j+k-i\). 

We have neglected the first and second derivatives of the amplitude \(A_i(z,t)\), due to the smallness of the \(\chi^{(3)}\).

\section{Simple coupled mode equation}\label{app:coupled_mode_equation}

\begin{figure}[htbp!]
\includegraphics[width=7cm]{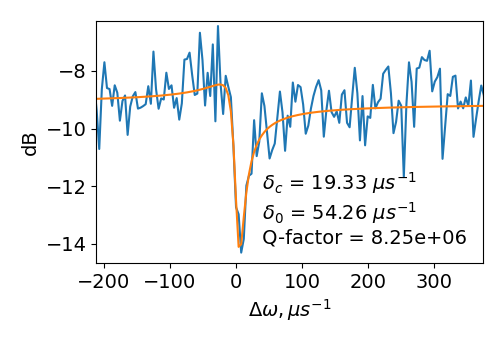}
\caption{An example of measured transmission spectrum with the Fano profile approximation.}\label{Fig:trans_spectrum}
\end{figure}

Within the simple coupling model \cite{Gorodetsky:99}, the equation for a slowly varying mode amplitude in a linear regime with a spatial distribution \(e_m(\vec r)\) in a microcavity, where the field is defined as \(E=(a(t)e_m(\vec r)e ^{i\omega_p t}+c.c.)/2\), \(m\) - the mode number, \(\omega_p\) - the pump frequency\cite{Gorodetsky:99}:

\begin{equation}
     i\cfrac{\partial a(t)}{\partial t}-\Delta\omega a(t)+(\delta_0+\delta_c)a(t)=iF,\label{Eq_simple_coupled_model}
\end{equation}

\(F=\sqrt{4P_{in}\delta_c/(\varepsilon_0
\varepsilon V_{ef})}\), where \(P_{in}\) - the pump power, \(V_{ef}=\int |e_m(\vec r)|^2d^3r\) - the effective mode volume, \(\max e_m(\vec r)=1\), $\delta_c$ is the coupling strength coefficient, $\delta_0$ is the losses experienced by the whispering gallery modes,  $\Delta\omega_q=\omega_p-\omega_{m,q}$ is the pump frequency tuning.
Stationary solution of the equation \eqref{Eq_simple_coupled_model} leads to the transmission spectrum $T$ of the microresonator-taper system, that is an experimentally determined quantity, allowing  to determine the parameters $\delta_c, \delta_0$. Thus, in the case of a single-mode coupling element, the transmission spectrum is described by the Fano resonance profile\cite{limonov2017fano, Sumetsky:12, lu2019tunable}:
\begin{equation}
    {|T|^2}=|S_0|^2\left|e^{i\varphi_0}-\cfrac{2\delta_c}{i\Delta\omega_q+(\delta_0+\delta_c)}\right|^2.\label{Eq_Fano_shape}
\end{equation}
Here \(S_0 = |S_0|e^{i\varphi_0}\) is the nonresonant transmission coefficient,

Figure \ref{Fig:trans_spectrum} shows the example of such a transmission spectrum, measured for a particular axial-azimuthal mode in a SNAP microresonator with approximation with \eqref{Eq_Fano_shape}.

Approximation of the transmission spectrum makes it possible to determine the coupling parameters \(\delta_0\) and \(\delta_c\), which contain information about the overlap integrals of the radiation source field with the resonator mode and can be expressed in terms of the required parameters \(D_m\) and \( C_m\).
\bibliography{main.bib}

\begin{thebibliography}{22}%
\makeatletter
\providecommand \@ifxundefined [1]{%
 \@ifx{#1\undefined}
}%
\providecommand \@ifnum [1]{%
 \ifnum #1\expandafter \@firstoftwo
 \else \expandafter \@secondoftwo
 \fi
}%
\providecommand \@ifx [1]{%
 \ifx #1\expandafter \@firstoftwo
 \else \expandafter \@secondoftwo
 \fi
}%
\providecommand \natexlab [1]{#1}%
\providecommand \enquote  [1]{``#1''}%
\providecommand \bibnamefont  [1]{#1}%
\providecommand \bibfnamefont [1]{#1}%
\providecommand \citenamefont [1]{#1}%
\providecommand \href@noop [0]{\@secondoftwo}%
\providecommand \href [0]{\begingroup \@sanitize@url \@href}%
\providecommand \@href[1]{\@@startlink{#1}\@@href}%
\providecommand \@@href[1]{\endgroup#1\@@endlink}%
\providecommand \@sanitize@url [0]{\catcode `\\12\catcode `\$12\catcode
  `\&12\catcode `\#12\catcode `\^12\catcode `\_12\catcode `\%12\relax}%
\providecommand \@@startlink[1]{}%
\providecommand \@@endlink[0]{}%
\providecommand \url  [0]{\begingroup\@sanitize@url \@url }%
\providecommand \@url [1]{\endgroup\@href {#1}{\urlprefix }}%
\providecommand \urlprefix  [0]{URL }%
\providecommand \Eprint [0]{\href }%
\providecommand \doibase [0]{https://doi.org/}%
\providecommand \selectlanguage [0]{\@gobble}%
\providecommand \bibinfo  [0]{\@secondoftwo}%
\providecommand \bibfield  [0]{\@secondoftwo}%
\providecommand \translation [1]{[#1]}%
\providecommand \BibitemOpen [0]{}%
\providecommand \bibitemStop [0]{}%
\providecommand \bibitemNoStop [0]{.\EOS\space}%
\providecommand \EOS [0]{\spacefactor3000\relax}%
\providecommand \BibitemShut  [1]{\csname bibitem#1\endcsname}%
\let\auto@bib@innerbib\@empty
\bibitem [{\citenamefont {Pasquazi}\ \emph {et~al.}(2018)\citenamefont
  {Pasquazi}, \citenamefont {Peccianti}, \citenamefont {Razzari}, \citenamefont
  {Moss}, \citenamefont {Coen}, \citenamefont {Erkintalo}, \citenamefont
  {Chembo}, \citenamefont {Hansson}, \citenamefont {Wabnitz}, \citenamefont
  {Del’Haye}, \citenamefont {Xue}, \citenamefont {Weiner},\ and\
  \citenamefont {Morandotti}}]{PASQUAZI20181}%
  \BibitemOpen
  \bibfield  {author} {\bibinfo {author} {\bibfnamefont {A.}~\bibnamefont
  {Pasquazi}}, \bibinfo {author} {\bibfnamefont {M.}~\bibnamefont {Peccianti}},
  \bibinfo {author} {\bibfnamefont {L.}~\bibnamefont {Razzari}}, \bibinfo
  {author} {\bibfnamefont {D.~J.}\ \bibnamefont {Moss}}, \bibinfo {author}
  {\bibfnamefont {S.}~\bibnamefont {Coen}}, \bibinfo {author} {\bibfnamefont
  {M.}~\bibnamefont {Erkintalo}}, \bibinfo {author} {\bibfnamefont {Y.~K.}\
  \bibnamefont {Chembo}}, \bibinfo {author} {\bibfnamefont {T.}~\bibnamefont
  {Hansson}}, \bibinfo {author} {\bibfnamefont {S.}~\bibnamefont {Wabnitz}},
  \bibinfo {author} {\bibfnamefont {P.}~\bibnamefont {Del’Haye}}, \bibinfo
  {author} {\bibfnamefont {X.}~\bibnamefont {Xue}}, \bibinfo {author}
  {\bibfnamefont {A.~M.}\ \bibnamefont {Weiner}},\ and\ \bibinfo {author}
  {\bibfnamefont {R.}~\bibnamefont {Morandotti}},\ }\bibfield  {title}
  {\bibinfo {title} {Micro-combs: A novel generation of optical sources},\
  }\href {https://doi.org/https://doi.org/10.1016/j.physrep.2017.08.004}
  {\bibfield  {journal} {\bibinfo  {journal} {Physics Reports}\ }\textbf
  {\bibinfo {volume} {729}},\ \bibinfo {pages} {1} (\bibinfo {year}
  {2018})}\BibitemShut {NoStop}%
\bibitem [{\citenamefont {Niu}\ \emph {et~al.}(2023)\citenamefont {Niu},
  \citenamefont {Li}, \citenamefont {Wan}, \citenamefont {Sun}, \citenamefont
  {Hu}, \citenamefont {Zou}, \citenamefont {Guo},\ and\ \citenamefont
  {Dong}}]{Niu2023}%
  \BibitemOpen
  \bibfield  {author} {\bibinfo {author} {\bibfnamefont {R.}~\bibnamefont
  {Niu}}, \bibinfo {author} {\bibfnamefont {M.}~\bibnamefont {Li}}, \bibinfo
  {author} {\bibfnamefont {S.}~\bibnamefont {Wan}}, \bibinfo {author}
  {\bibfnamefont {Y.~R.}\ \bibnamefont {Sun}}, \bibinfo {author} {\bibfnamefont
  {S.-m.}\ \bibnamefont {Hu}}, \bibinfo {author} {\bibfnamefont {C.-l.}\
  \bibnamefont {Zou}}, \bibinfo {author} {\bibfnamefont {G.-c.}\ \bibnamefont
  {Guo}},\ and\ \bibinfo {author} {\bibfnamefont {C.-h.}\ \bibnamefont
  {Dong}},\ }\bibfield  {title} {\bibinfo {title} {{kHz-precision wavemeter
  based on reconfigurable microsoliton}},\ }\href
  {https://doi.org/10.1038/s41467-022-35728-x} {\bibfield  {journal} {\bibinfo
  {journal} {Nature Communications}\ }\textbf {\bibinfo {volume} {14}},\
  \bibinfo {pages} {169} (\bibinfo {year} {2023})}\BibitemShut {NoStop}%
\bibitem [{\citenamefont {Sugiyama}\ \emph {et~al.}(2023)\citenamefont
  {Sugiyama}, \citenamefont {Kashimura}, \citenamefont {Kashimoto},
  \citenamefont {Akamatsu},\ and\ \citenamefont {Hong}}]{Sugiyama2023}%
  \BibitemOpen
  \bibfield  {author} {\bibinfo {author} {\bibfnamefont {Y.}~\bibnamefont
  {Sugiyama}}, \bibinfo {author} {\bibfnamefont {T.}~\bibnamefont {Kashimura}},
  \bibinfo {author} {\bibfnamefont {K.}~\bibnamefont {Kashimoto}}, \bibinfo
  {author} {\bibfnamefont {D.}~\bibnamefont {Akamatsu}},\ and\ \bibinfo
  {author} {\bibfnamefont {F.-L.}\ \bibnamefont {Hong}},\ }\bibfield  {title}
  {\bibinfo {title} {{Precision dual-comb spectroscopy using
  wavelength-converted frequency combs with low repetition rates}},\ }\href
  {https://doi.org/10.1038/s41598-023-29734-2} {\bibfield  {journal} {\bibinfo
  {journal} {Scientific Reports}\ }\textbf {\bibinfo {volume} {13}},\ \bibinfo
  {pages} {2549} (\bibinfo {year} {2023})}\BibitemShut {NoStop}%
\bibitem [{\citenamefont {Sumetsky}\ and\ \citenamefont
  {Fini}(2011)}]{Sumetsky:11}%
  \BibitemOpen
  \bibfield  {author} {\bibinfo {author} {\bibfnamefont {M.}~\bibnamefont
  {Sumetsky}}\ and\ \bibinfo {author} {\bibfnamefont {J.~M.}\ \bibnamefont
  {Fini}},\ }\bibfield  {title} {\bibinfo {title} {Surface nanoscale axial
  photonics},\ }\href {https://doi.org/10.1364/OE.19.026470} {\bibfield
  {journal} {\bibinfo  {journal} {Opt. Express}\ }\textbf {\bibinfo {volume}
  {19}},\ \bibinfo {pages} {26470} (\bibinfo {year} {2011})}\BibitemShut
  {NoStop}%
\bibitem [{\citenamefont {Bochek}\ \emph {et~al.}(2019)\citenamefont {Bochek},
  \citenamefont {Toropov}, \citenamefont {Vatnik}, \citenamefont {Churkin},\
  and\ \citenamefont {Sumetsky}}]{Bochek:19}%
  \BibitemOpen
  \bibfield  {author} {\bibinfo {author} {\bibfnamefont {D.}~\bibnamefont
  {Bochek}}, \bibinfo {author} {\bibfnamefont {N.}~\bibnamefont {Toropov}},
  \bibinfo {author} {\bibfnamefont {I.}~\bibnamefont {Vatnik}}, \bibinfo
  {author} {\bibfnamefont {D.}~\bibnamefont {Churkin}},\ and\ \bibinfo {author}
  {\bibfnamefont {M.}~\bibnamefont {Sumetsky}},\ }\bibfield  {title} {\bibinfo
  {title} {Snap microresonators introduced by strong bending of optical
  fibers},\ }\href {https://doi.org/10.1364/OL.44.003218} {\bibfield  {journal}
  {\bibinfo  {journal} {Opt. Lett.}\ }\textbf {\bibinfo {volume} {44}},\
  \bibinfo {pages} {3218} (\bibinfo {year} {2019})}\BibitemShut {NoStop}%
\bibitem [{\citenamefont {Jin}\ \emph {et~al.}(2021)\citenamefont {Jin},
  \citenamefont {Xu}, \citenamefont {Gao}, \citenamefont {Wang}, \citenamefont
  {Xia},\ and\ \citenamefont {Yu}}]{Jin:21}%
  \BibitemOpen
  \bibfield  {author} {\bibinfo {author} {\bibfnamefont {X.}~\bibnamefont
  {Jin}}, \bibinfo {author} {\bibfnamefont {X.}~\bibnamefont {Xu}}, \bibinfo
  {author} {\bibfnamefont {H.}~\bibnamefont {Gao}}, \bibinfo {author}
  {\bibfnamefont {K.}~\bibnamefont {Wang}}, \bibinfo {author} {\bibfnamefont
  {H.}~\bibnamefont {Xia}},\ and\ \bibinfo {author} {\bibfnamefont
  {L.}~\bibnamefont {Yu}},\ }\bibfield  {title} {\bibinfo {title} {Controllable
  two-dimensional kerr and raman-kerr frequency combs in microbottle resonators
  with selectable dispersion},\ }\href {https://doi.org/10.1364/PRJ.408492}
  {\bibfield  {journal} {\bibinfo  {journal} {Photon. Res.}\ }\textbf {\bibinfo
  {volume} {9}},\ \bibinfo {pages} {171} (\bibinfo {year} {2021})}\BibitemShut
  {NoStop}%
\bibitem [{\citenamefont {P\"{o}llinger}\ and\ \citenamefont
  {Rauschenbeutel}(2010)}]{Pollinger:10}%
  \BibitemOpen
  \bibfield  {author} {\bibinfo {author} {\bibfnamefont {M.}~\bibnamefont
  {P\"{o}llinger}}\ and\ \bibinfo {author} {\bibfnamefont {A.}~\bibnamefont
  {Rauschenbeutel}},\ }\bibfield  {title} {\bibinfo {title} {All-optical signal
  processing at ultra-low powers in bottle microresonators using the kerr
  effect},\ }\href {https://doi.org/10.1364/OE.18.017764} {\bibfield  {journal}
  {\bibinfo  {journal} {Opt. Express}\ }\textbf {\bibinfo {volume} {18}},\
  \bibinfo {pages} {17764} (\bibinfo {year} {2010})}\BibitemShut {NoStop}%
\bibitem [{\citenamefont {Zhu}\ \emph {et~al.}(2019)\citenamefont {Zhu},
  \citenamefont {Xiao}, \citenamefont {Jiang}, \citenamefont {Shi},\ and\
  \citenamefont {Zhang}}]{ZhuXiaoJiangShiZhang+2019+931+940}%
  \BibitemOpen
  \bibfield  {author} {\bibinfo {author} {\bibfnamefont {S.}~\bibnamefont
  {Zhu}}, \bibinfo {author} {\bibfnamefont {B.}~\bibnamefont {Xiao}}, \bibinfo
  {author} {\bibfnamefont {B.}~\bibnamefont {Jiang}}, \bibinfo {author}
  {\bibfnamefont {L.}~\bibnamefont {Shi}},\ and\ \bibinfo {author}
  {\bibfnamefont {X.}~\bibnamefont {Zhang}},\ }\bibfield  {title} {\bibinfo
  {title} {Tunable brillouin and raman microlasers using hybrid microbottle
  resonators},\ }\href {https://doi.org/doi:10.1515/nanoph-2019-0070}
  {\bibfield  {journal} {\bibinfo  {journal} {Nanophotonics}\ }\textbf
  {\bibinfo {volume} {8}},\ \bibinfo {pages} {931} (\bibinfo {year}
  {2019})}\BibitemShut {NoStop}%
\bibitem [{\citenamefont {Lugiato}\ \emph {et~al.}(2018)\citenamefont
  {Lugiato}, \citenamefont {Prati}, \citenamefont {Gorodetsky},\ and\
  \citenamefont {Kippenberg}}]{lugiato2018lugiato}%
  \BibitemOpen
  \bibfield  {author} {\bibinfo {author} {\bibfnamefont {L.}~\bibnamefont
  {Lugiato}}, \bibinfo {author} {\bibfnamefont {F.}~\bibnamefont {Prati}},
  \bibinfo {author} {\bibfnamefont {M.}~\bibnamefont {Gorodetsky}},\ and\
  \bibinfo {author} {\bibfnamefont {T.}~\bibnamefont {Kippenberg}},\ }\bibfield
   {title} {\bibinfo {title} {From the lugiato--lefever equation to
  microresonator-based soliton kerr frequency combs},\ }\href@noop {}
  {\bibfield  {journal} {\bibinfo  {journal} {Philosophical Transactions of the
  Royal Society A: Mathematical, Physical and Engineering Sciences}\ }\textbf
  {\bibinfo {volume} {376}},\ \bibinfo {pages} {20180113} (\bibinfo {year}
  {2018})}\BibitemShut {NoStop}%
\bibitem [{\citenamefont {Kartashov}\ \emph {et~al.}(2018)\citenamefont
  {Kartashov}, \citenamefont {Gorodetsky}, \citenamefont {Kudlinski},\ and\
  \citenamefont {Skryabin}}]{Kartashov:18}%
  \BibitemOpen
  \bibfield  {author} {\bibinfo {author} {\bibfnamefont {Y.~V.}\ \bibnamefont
  {Kartashov}}, \bibinfo {author} {\bibfnamefont {M.~L.}\ \bibnamefont
  {Gorodetsky}}, \bibinfo {author} {\bibfnamefont {A.}~\bibnamefont
  {Kudlinski}},\ and\ \bibinfo {author} {\bibfnamefont {D.~V.}\ \bibnamefont
  {Skryabin}},\ }\bibfield  {title} {\bibinfo {title} {Two-dimensional
  nonlinear modes and frequency combs in bottle microresonators},\ }\href
  {https://doi.org/10.1364/OL.43.002680} {\bibfield  {journal} {\bibinfo
  {journal} {Opt. Lett.}\ }\textbf {\bibinfo {volume} {43}},\ \bibinfo {pages}
  {2680} (\bibinfo {year} {2018})}\BibitemShut {NoStop}%
\bibitem [{\citenamefont {Oreshnikov}\ and\ \citenamefont
  {Skryabin}(2017)}]{Oreshnikov:17}%
  \BibitemOpen
  \bibfield  {author} {\bibinfo {author} {\bibfnamefont {I.}~\bibnamefont
  {Oreshnikov}}\ and\ \bibinfo {author} {\bibfnamefont {D.~V.}\ \bibnamefont
  {Skryabin}},\ }\bibfield  {title} {\bibinfo {title} {Multiple nonlinear
  resonances and frequency combs in bottle microresonators},\ }\href
  {https://doi.org/10.1364/OE.25.010306} {\bibfield  {journal} {\bibinfo
  {journal} {Opt. Express}\ }\textbf {\bibinfo {volume} {25}},\ \bibinfo
  {pages} {10306} (\bibinfo {year} {2017})}\BibitemShut {NoStop}%
\bibitem [{\citenamefont {Sumetsky}(2012)}]{Sumetsky:12}%
  \BibitemOpen
  \bibfield  {author} {\bibinfo {author} {\bibfnamefont {M.}~\bibnamefont
  {Sumetsky}},\ }\bibfield  {title} {\bibinfo {title} {Theory of {SNAP}
  devices: basic equations and comparison with the experiment},\ }\href
  {https://doi.org/10.1364/OE.20.022537} {\bibfield  {journal} {\bibinfo
  {journal} {Opt. Express}\ }\textbf {\bibinfo {volume} {20}},\ \bibinfo
  {pages} {22537} (\bibinfo {year} {2012})}\BibitemShut {NoStop}%
\bibitem [{\citenamefont {Suchkov}\ \emph {et~al.}(2017)\citenamefont
  {Suchkov}, \citenamefont {Sumetsky},\ and\ \citenamefont
  {Sukhorukov}}]{Suchkov:17}%
  \BibitemOpen
  \bibfield  {author} {\bibinfo {author} {\bibfnamefont {S.~V.}\ \bibnamefont
  {Suchkov}}, \bibinfo {author} {\bibfnamefont {M.}~\bibnamefont {Sumetsky}},\
  and\ \bibinfo {author} {\bibfnamefont {A.~A.}\ \bibnamefont {Sukhorukov}},\
  }\bibfield  {title} {\bibinfo {title} {Frequency comb generation in snap
  bottle resonators},\ }\href {https://doi.org/10.1364/OL.42.002149} {\bibfield
   {journal} {\bibinfo  {journal} {Opt. Lett.}\ }\textbf {\bibinfo {volume}
  {42}},\ \bibinfo {pages} {2149} (\bibinfo {year} {2017})}\BibitemShut
  {NoStop}%
\bibitem [{\citenamefont {Crespo-Ballesteros}\ \emph
  {et~al.}(2022)\citenamefont {Crespo-Ballesteros}, \citenamefont {Matsko},\
  and\ \citenamefont {Sumetsky}}]{Crespo-Ballesteros2022}%
  \BibitemOpen
  \bibfield  {author} {\bibinfo {author} {\bibfnamefont {M.}~\bibnamefont
  {Crespo-Ballesteros}}, \bibinfo {author} {\bibfnamefont {A.}~\bibnamefont
  {Matsko}},\ and\ \bibinfo {author} {\bibfnamefont {M.}~\bibnamefont
  {Sumetsky}},\ }\bibfield  {title} {\bibinfo {title} {{Optimized frequency
  comb spectrum of parametrically modulated bottle microresonators}},\
  }\bibfield  {journal} {\bibinfo  {journal} {arXiv}\ }\href
  {https://doi.org/10.48550} {10.48550} (\bibinfo {year} {2022}),\ \Eprint
  {https://arxiv.org/abs/2211.01349} {arXiv:2211.01349} \BibitemShut {NoStop}%
\bibitem [{\citenamefont {Crespo-Ballesteros}\ and\ \citenamefont
  {Sumetsky}(2021)}]{Crespo-Ballesteros2021}%
  \BibitemOpen
  \bibfield  {author} {\bibinfo {author} {\bibfnamefont {M.}~\bibnamefont
  {Crespo-Ballesteros}}\ and\ \bibinfo {author} {\bibfnamefont
  {M.}~\bibnamefont {Sumetsky}},\ }\bibfield  {title} {\bibinfo {title}
  {{Controlled Transportation of Light by Light at the Microscale}},\ }\href
  {https://doi.org/10.1103/PhysRevLett.126.153901} {\bibfield  {journal}
  {\bibinfo  {journal} {Physical Review Letters}\ }\textbf {\bibinfo {volume}
  {126}},\ \bibinfo {pages} {153901} (\bibinfo {year} {2021})}\BibitemShut
  {NoStop}%
\bibitem [{\citenamefont {Kolesnikova}\ \emph {et~al.}(2022)\citenamefont
  {Kolesnikova}, \citenamefont {Suchkov},\ and\ \citenamefont
  {Vatnik}}]{Kolesnikova:22}%
  \BibitemOpen
  \bibfield  {author} {\bibinfo {author} {\bibfnamefont {A.~Y.}\ \bibnamefont
  {Kolesnikova}}, \bibinfo {author} {\bibfnamefont {S.~V.}\ \bibnamefont
  {Suchkov}},\ and\ \bibinfo {author} {\bibfnamefont {I.~D.}\ \bibnamefont
  {Vatnik}},\ }\bibfield  {title} {\bibinfo {title} {Frequency comb generation
  in snap fiber resonator based on axial-azimuthal mode interactions},\ }\href
  {https://doi.org/10.1364/OE.450298} {\bibfield  {journal} {\bibinfo
  {journal} {Opt. Express}\ }\textbf {\bibinfo {volume} {30}},\ \bibinfo
  {pages} {10588} (\bibinfo {year} {2022})}\BibitemShut {NoStop}%
\bibitem [{\citenamefont {Demchenko}\ and\ \citenamefont
  {Gorodetsky}(2013)}]{Demchenko:13}%
  \BibitemOpen
  \bibfield  {author} {\bibinfo {author} {\bibfnamefont {Y.~A.}\ \bibnamefont
  {Demchenko}}\ and\ \bibinfo {author} {\bibfnamefont {M.~L.}\ \bibnamefont
  {Gorodetsky}},\ }\bibfield  {title} {\bibinfo {title} {Analytical estimates
  of eigenfrequencies, dispersion, and field distribution in whispering gallery
  resonators},\ }\href {https://doi.org/10.1364/JOSAB.30.003056} {\bibfield
  {journal} {\bibinfo  {journal} {J. Opt. Soc. Am. B}\ }\textbf {\bibinfo
  {volume} {30}},\ \bibinfo {pages} {3056} (\bibinfo {year}
  {2013})}\BibitemShut {NoStop}%
\bibitem [{\citenamefont {Vitullo}\ \emph {et~al.}(2020)\citenamefont
  {Vitullo}, \citenamefont {Zaki}, \citenamefont {Jones}, \citenamefont
  {Sumetsky},\ and\ \citenamefont {Brodsky}}]{Vitullo:20}%
  \BibitemOpen
  \bibfield  {author} {\bibinfo {author} {\bibfnamefont {D.~L.~P.}\
  \bibnamefont {Vitullo}}, \bibinfo {author} {\bibfnamefont {S.}~\bibnamefont
  {Zaki}}, \bibinfo {author} {\bibfnamefont {D.~E.}\ \bibnamefont {Jones}},
  \bibinfo {author} {\bibfnamefont {M.}~\bibnamefont {Sumetsky}},\ and\
  \bibinfo {author} {\bibfnamefont {M.}~\bibnamefont {Brodsky}},\ }\bibfield
  {title} {\bibinfo {title} {Coupling between waveguides and microresonators:
  the local approach},\ }\href {https://doi.org/10.1364/OE.399978} {\bibfield
  {journal} {\bibinfo  {journal} {Opt. Express}\ }\textbf {\bibinfo {volume}
  {28}},\ \bibinfo {pages} {25908} (\bibinfo {year} {2020})}\BibitemShut
  {NoStop}%
\bibitem [{\citenamefont {Herr}\ \emph {et~al.}(2012)\citenamefont {Herr},
  \citenamefont {Hartinger}, \citenamefont {Riemensberger}, \citenamefont
  {Wang}, \citenamefont {Gavartin}, \citenamefont {Holzwarth}, \citenamefont
  {Gorodetsky},\ and\ \citenamefont {Kippenberg}}]{Herr2012}%
  \BibitemOpen
  \bibfield  {author} {\bibinfo {author} {\bibfnamefont {T.}~\bibnamefont
  {Herr}}, \bibinfo {author} {\bibfnamefont {K.}~\bibnamefont {Hartinger}},
  \bibinfo {author} {\bibfnamefont {J.}~\bibnamefont {Riemensberger}}, \bibinfo
  {author} {\bibfnamefont {C.~Y.}\ \bibnamefont {Wang}}, \bibinfo {author}
  {\bibfnamefont {E.}~\bibnamefont {Gavartin}}, \bibinfo {author}
  {\bibfnamefont {R.}~\bibnamefont {Holzwarth}}, \bibinfo {author}
  {\bibfnamefont {M.~L.}\ \bibnamefont {Gorodetsky}},\ and\ \bibinfo {author}
  {\bibfnamefont {T.~J.}\ \bibnamefont {Kippenberg}},\ }\bibfield  {title}
  {\bibinfo {title} {{Universal dynamics of kerr-frequency comb formation in
  microresonators}},\ }\href {https://doi.org/10.1038/nphoton.2012.127}
  {\bibfield  {journal} {\bibinfo  {journal} {Optics InfoBase Conference
  Papers}\ ,\ \bibinfo {pages} {1}} (\bibinfo {year} {2012})},\ \Eprint
  {https://arxiv.org/abs/1111.3071} {arXiv:1111.3071} \BibitemShut {NoStop}%
\bibitem [{\citenamefont {Gorodetsky}\ and\ \citenamefont
  {Ilchenko}(1999)}]{Gorodetsky:99}%
  \BibitemOpen
  \bibfield  {author} {\bibinfo {author} {\bibfnamefont {M.~L.}\ \bibnamefont
  {Gorodetsky}}\ and\ \bibinfo {author} {\bibfnamefont {V.~S.}\ \bibnamefont
  {Ilchenko}},\ }\bibfield  {title} {\bibinfo {title} {Optical microsphere
  resonators: optimal coupling to high-q whispering-gallery modes},\ }\href
  {https://doi.org/10.1364/JOSAB.16.000147} {\bibfield  {journal} {\bibinfo
  {journal} {J. Opt. Soc. Am. B}\ }\textbf {\bibinfo {volume} {16}},\ \bibinfo
  {pages} {147} (\bibinfo {year} {1999})}\BibitemShut {NoStop}%
\bibitem [{\citenamefont {Limonov}\ \emph {et~al.}(2017)\citenamefont
  {Limonov}, \citenamefont {Rybin}, \citenamefont {Poddubny},\ and\
  \citenamefont {Kivshar}}]{limonov2017fano}%
  \BibitemOpen
  \bibfield  {author} {\bibinfo {author} {\bibfnamefont {M.~F.}\ \bibnamefont
  {Limonov}}, \bibinfo {author} {\bibfnamefont {M.~V.}\ \bibnamefont {Rybin}},
  \bibinfo {author} {\bibfnamefont {A.~N.}\ \bibnamefont {Poddubny}},\ and\
  \bibinfo {author} {\bibfnamefont {Y.~S.}\ \bibnamefont {Kivshar}},\
  }\bibfield  {title} {\bibinfo {title} {Fano resonances in photonics},\
  }\href@noop {} {\bibfield  {journal} {\bibinfo  {journal} {Nature Photonics}\
  }\textbf {\bibinfo {volume} {11}},\ \bibinfo {pages} {543} (\bibinfo {year}
  {2017})}\BibitemShut {NoStop}%
\bibitem [{\citenamefont {Lu}\ \emph {et~al.}(2019)\citenamefont {Lu},
  \citenamefont {Zhu}, \citenamefont {Li}, \citenamefont {Nie}, \citenamefont
  {Li},\ and\ \citenamefont {Song}}]{lu2019tunable}%
  \BibitemOpen
  \bibfield  {author} {\bibinfo {author} {\bibfnamefont {Y.}~\bibnamefont
  {Lu}}, \bibinfo {author} {\bibfnamefont {X.}~\bibnamefont {Zhu}}, \bibinfo
  {author} {\bibfnamefont {J.}~\bibnamefont {Li}}, \bibinfo {author}
  {\bibfnamefont {Y.}~\bibnamefont {Nie}}, \bibinfo {author} {\bibfnamefont
  {M.}~\bibnamefont {Li}},\ and\ \bibinfo {author} {\bibfnamefont
  {Y.}~\bibnamefont {Song}},\ }\bibfield  {title} {\bibinfo {title} {Tunable
  oscillating fano spectra in a fiber taper coupled conical microresonator},\
  }\href@noop {} {\bibfield  {journal} {\bibinfo  {journal} {IEEE Photonics
  Journal}\ }\textbf {\bibinfo {volume} {11}},\ \bibinfo {pages} {1} (\bibinfo
  {year} {2019})}\BibitemShut {NoStop}%
\end{thebibliography}%

\end{document}